\documentclass[prl,twocolumn,aps]{revtex4}

\usepackage{graphicx}
\usepackage{dcolumn}
\usepackage{bm}
\usepackage[dvips]{color}

\def\mrm{\mathrm}

\def\mrm{\mathrm}

\def\del{\partial}

\def\delij{\delta_{\sigma_i \sigma_j}}
\def\deli0{\delta_{\sigma_i 0}}
\def\delj0{\delta_{\sigma_j 0}}

\definecolor{darkgreen}{rgb}{0.0, 0.5, 0.0}

\def\n+{_{n+1}}
\def\sn{_\mrm{sn}}

\def\Ell{J}
\renewcommand{\ell}{j}

\begin{document}


\preprint{APS/123-QED}

\title{
Generalized scaling theory for critical phenomena \\
including essential singularity and infinite dimensionality
}

\author{Tomoaki Nogawa}
\email{nogawa@serow.t.u-tokyo.ac.jp}
\affiliation{%
Department of Applied Physics, 
The University of Tokyo, 7-3-1, Hongo, Bunkyo-ku, Tokyo 113-8656, Japan
}
\author{Takehisa Hasegawa}
\affiliation{%
Graduate School of Information Science, 
Tohoku University, 
6-3-09, Aramaki-Aza-Aoba, Sendai, Miyagi 980-8579, Japan
}%
\author{Koji Nemoto}
\affiliation{%
Department of Physics, Hokkaido University,
Kita 10 Nisi 8, Kita-ku
Sapporo, Hokkaido 060-0810, Japan
}%

\begin{abstract}
We propose a generic scaling theory for critical phenomena 
that includes power-law and essential singularities 
in finite and infinite dimensional systems. 
In addition, we clarify its validity by analyzing the Potts model 
in a simple hierarchical network, 
where a saddle-node bifurcation of the renormalization-group fixed point 
governs the essential singularity.
\end{abstract}



\maketitle


The scaling theory for the power-law singularity (PLS) 
of second-order transitions is among the most significant achievements 
in theoretical physics \cite{Wilson75}, 
and provides a comprehensive understanding 
of critical phenomena with minimal assumptions. 
The key concept of this theory is an invariance of the singular part of free energy, $g_d$, 
which is scaled by a factor $\hat{b}$ as
\begin{equation}
g_d(t,h,L^{-1}) = \hat{b}^{-d} g_d(  t \hat{b}^{y_t}, h \hat{b}^{y_h},  L^{-1} \hat{b} ). 
\label{eq:ordinary}
\end{equation}
Here, $d$ is the dimension of space, $L$ is the linear dimension 
of the system, and $t$ and $h$ are deviations of model parameters 
from their critical values 
(such as reduced temperature $t=T-T_c$ and magnetic field  $h=H$). 
For the characterization of the criticality, 
$y_t$ and $y_h$ are the most fundamental quantities, 
which lead to critical exponents for the PLS 
of physical quantities 
e.g.,  $\nu=1/y_t$ and $\beta=(d-y_h)/y_t$, 
corresponding to the correlation length as $\hat{\xi} \propto t^{-\nu}$ 
and the order parameter as $m \propto t^\beta$. 
The scaling theory is based on the self-similarity 
owing to the divergence of the correlation length.


There is a kind of continuous transition 
on which standard scaling theory does not work. 
The most famous example is the Berezinskii-Kosterlitz-Thouless (BKT) 
transition in the two-imensional XY model 
\cite{Berezinskii72, Kosterlitz73, Kosterlitz74}, 
which exhibits an essential singularity (ES) around the transition point 
in the disordered phase, where the correlation length diverges as 
$\hat{\xi} \propto \exp[ (t/t_0)^{-1/2} ]$ instead of the power-law. 
Below the transition temperature, $\tilde \xi$ remains divergent and
there appears quasi-long-range order with zero magnetization 
as if the system stays in criticality.
In this sense we hereafter call this phase the {\em critical phase}.

A similar singularity has been recently found in infinite dimensional systems. 
The ES for the infinite dimensional system appears in the {\it ordered} phase; 
the order parameter behaves as $m \propto \exp[ -(t/t_0)^{-1/2} ]$ 
toward the transition point, above which the critical phase emerges.
This is in contrast to the ordinary BKT transition, 
and is called the inverted BKT singularity \cite{Hinczewski06}. 
Here we use the word ``infinite dimensional'' for the property 
that the typical path length $L$ increases with system volume $N$ 
as $L \propto \log N$ instead of $N^{1/d}$. 
This property is realized in trees, random graphs \cite{Newman01}, 
hierarchical lattices \cite{Hinczewski06}, hyperbolic lattices \cite{Shima06}, 
and small-world networks \cite{Watts98} among others. 
Such systems have been extensively studied in the context of complex networks; 
the heterogeneous and hierarchical structure has unexpectedly revealed 
important concepts in physics \cite{Dorogovtsev08} 
other than the mean-field behaviors for simpler infinite dimensional systems 
such as the complete graph. 
In particular, the ES is found rather often in various models 
on various infinite dimensional graphs 
\cite{Krapivsky04, Bauer05, Bollobas05, Riordan05, Boettcher09, 
Berker09, Hasegawa10, Hinczewski06, Boettcher11}. 
Therefore, the ES is considered to be a basic concepts of the dynamics 
in infinite dimensional systems rather than an exotic topic 
as the BKT transition in finite dimensions. 
Since inverted BKT singularity is observed in percolation 
\cite{Krapivsky04, Bauer05, Bollobas05, Riordan05, Boettcher09, Berker09, Hasegawa10} 
and Ising models \cite{Hinczewski06, Boettcher11}, 
we do not expect its mechanism to be common to that of the ordinary BKT transition, 
such as vortex-pair condensation \cite{Berezinskii72, Kosterlitz73}. 
Universal understanding of the ES in infinite dimensional systems is still missing.


In this Letter, we propose a generic scaling theory 
for critical phenomena with ES in infinite dimensional graphs. 
We also perform renormalization group (RG) analysis 
of the Potts model in a simple hierarchical network, 
which confirms the validity of the scaling law.



Now, we consider a scaling formula for infinite dimensional graphs as 
\begin{eqnarray}
g( \xi(t)^{-1}, h, N^{-1}) = b g( \xi(t)^{-1} b, h b^{y_h},  N^{-1} b ).
\label{eq:free_energy0}
\end{eqnarray}
Here, $b$ is a scaling factor for the total volume (mass) $N$ 
and the correlation volume $\xi$ of the system. 
The length-based expression is obtained by replacing  
$N \rightarrow L^d$, $b \rightarrow \hat{b}^d$, and $\xi(t) \rightarrow \hat{\xi}(t)^d$.
The volume-based expression can be used for the system where distance 
is not well defined, which is usual in infinite dimensional systems. 
Equation~(\ref{eq:free_energy0}) can be applied to various singularities 
by choosing a proper function form of $\xi(t)$. 
A similar idea is seen in the finite size scaling 
method proposed by Kim \cite{JKKim93, JKKim94} 
[it uses correlation length $\hat{\xi}(t)$ directly observed in advance].
If we assume $\xi$ is a power function as $t^{-1/y_t}$, 
Eq.~(\ref{eq:free_energy0}) leads to 
\begin{eqnarray}
g(t,h,N^{-1}) = b g(  t b^{y_t}, h b^{y_h},  N^{-1} b ), 
\label{eq:free_energy1}
\end{eqnarray}
which is essentially the same as Eq.~(\ref{eq:ordinary}).
On the other hand, by assuming an exponential function as 
$ 
\xi(t) = \exp[(t/t_0)^{-x_t}],
$ 
we obtain 
\begin{eqnarray}
g(t,h,N^{-1}) = b^{-1}
\hat{g}(  e^{-(t/t_0)^{-x_t}}b, h b^{y_h},  N^{-1} b ).
\label{eq:free_energy2}
\end{eqnarray}
Note that the present formula is applicable to normal phases, 
i.e., ordered or disordered phases, 
but not to critical phases where the correlation volume diverges.

We can calculate various physical quantities by differentiating the above free energy. 
Hereafter, we focus on the case of Eq.~(\ref{eq:free_energy2}) 
for the ES in the {\it ordered} phase.
The order parameter is given by 
\begin{eqnarray}
m(t,h,N^{-1}) = g_h(t,h,N^{-1}) 
\equiv \del g(t,h,N^{-1})/\del h
\nonumber \\
= b^{-(1-y_h)}
\hat{g}_h( e^{-(t/t_0)^{-x_t}} b, h b^{y_h}, N^{-1} b ).
\label{eq:order_parameter}
\end{eqnarray}

First, we consider the thermodynamic limit, $N=\infty$.
By setting $b = e^{(t/t_0)^{-x_t}}$, we have
\begin{equation}
m = e^{- (1-y_h) (t/t_0)^{-x_t} } 
\hat{g}_h( 1, h e^{y_h (t/t_0)^{-x_t}}, 0 ).
\end{equation}
Since $m$ is independent of $h$ for $h \rightarrow 0$ 
and independent of $t$ for $t \rightarrow 0$, 
the scaling function $g_h(1,x,0)$ should have asymptotic forms as 
\begin{equation}
\hat{g}_h(1,x,0) = \left \{
\begin{array}{ccc}
\mrm{const.} & \mrm{for} & x \ll 1
\\
x^{y_h^{-1}-1} & \mrm{for} & x \gg 1
\end{array}
\right. , 
\label{eq:scl1a}
\end{equation}
to reproduce 
\begin{equation}
m \propto \left\{
\begin{array}{ccc}
e^{-(1 - y_h) (t/t_0)^{-x_t} } & \mrm{for} & h e^{y_h (t/t_0)^{-x_t}} \ll 1
\\
h^{y_h^{-1}-1} & \mrm{for} & h e^{y_h (t/t_0)^{-x_t}} \gg 1
\end{array}
\right. .
\label{eq:scl1b}
\end{equation}
At $t=0$, it reads as $m\propto h^{1/\delta}$ with $1/\delta=y_h^{-1}-1$.

Second, we consider the case of $h=0$ with $N$ finite, where 
\begin{equation}
m = e^{-(1 - y_h) (t/t_0)^{-x_t} }
\hat{g}_h( 1, 0, N^{-1} e^{(t/t_0)^{-x_t}} ). 
\end{equation}
The scaling function should be 
\begin{equation}
\hat{g}_h(1,0,x) = \left \{
\begin{array}{ccc}
\mrm{const.}& \mrm{for} & x \ll 1
\\
x^{1-y_h} & \mrm{for} & x \gg 1
\end{array}
\right. ,
\end{equation}
in a similar manner as Eq.~(\ref{eq:scl1a}) to reproduce 
\begin{equation}
m \propto \left\{
\begin{array}{ccc}
e^{-(1-y_h) (t/t_0)^{-x_t} } & \mrm{for} & N^{-1} e^{(t/t_0)^{-x_t}} \ll 1
\\
N^{-(1-y_h)} & \mrm{for} & N^{-1} e^{(t/t_0)^{-x_t}} \gg 1
\end{array}
\right. .
\end{equation}
Similarly, susceptibility behaves as 
\begin{equation}
\chi \propto \left\{
\begin{array}{ccc}
e^{(2 y_h-1) (t/t_0)^{-x_t} } & \mrm{for} & N^{-1} e^{(t/t_0)^{-x_t}} \ll 1
\\
N^{2 y_h - 1} & \mrm{for} & N^{-1} e^{(t/t_0)^{-x_t}} \gg 1
\end{array}
\right. .
\end{equation} 
The $m$-th derivative of the free energy with $h$, 
which is proportional to $\xi(t)^{(m y_h - 1)}$, 
diverges at $t=0$ for $m > y_h^{-1}$. 
On the other hand, the derivative with $t$, such as specific heat, never diverges. 
Free energy for $h=0$ and $N=\infty$ is proportional to $\xi(t)^{-1}$, 
and the dominant term of its $m$-th derivative, $t^{-m(x_t+1)} \xi(t)^{-1}$, 
goes to zero for $t \rightarrow 0$.


Finally, we consider the case of $t=0$. 
By setting $b=N$ in Eq.~(\ref{eq:order_parameter}), we have 
\begin{eqnarray}
m = N^{-(1-y_h)} \hat{g}_h(0,h N^{y_h}, 1).
\end{eqnarray}
The scaling function should be 
\begin{equation}
\hat{g}_h(0,x,1) = \left \{
\begin{array}{ccc}
x & \mrm{for} & x \ll 1
\\
x^{y_h^{-1}-1} & \mrm{for} & x \gg 1
\end{array}
\right. , 
\label{eq:scl3a}
\end{equation}
to reproduce 
\begin{equation}
m \propto \left\{
\begin{array}{ccc}
h N^{2y_h-1} & \mrm{for} & h N^{y_h} \ll 1
\\
h^{y_h^{-1}-1} & \mrm{for} & h N^{y_h} \gg 1
\label{eq:scl3b}
\end{array}
\right. ,
\end{equation}
where we assumed linear susceptibility for a finite size system.
This form is the same as that of the conventional PLS.

\begin{figure}[t]
\begin{center}
\includegraphics[trim=250 110 410 50, scale=0.32, clip, angle=-0]{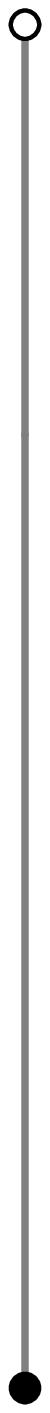}
\includegraphics[trim=250 110 410 50, scale=0.32, clip, angle=-0]{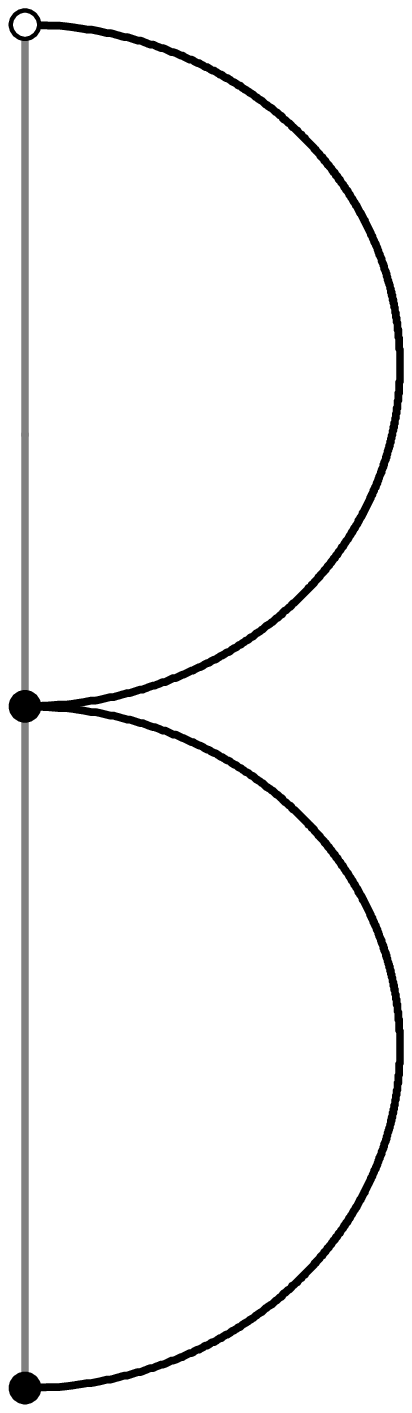}
\includegraphics[trim=250 110 410 50, scale=0.32, clip, angle=-0]{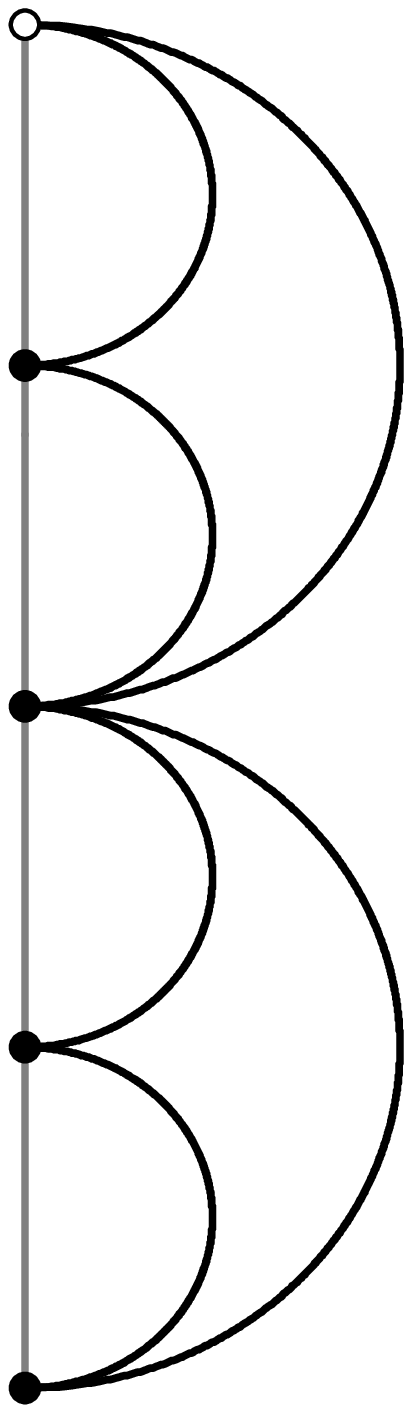}
\includegraphics[trim=250 110 410 50, scale=0.32, clip, angle=-0]{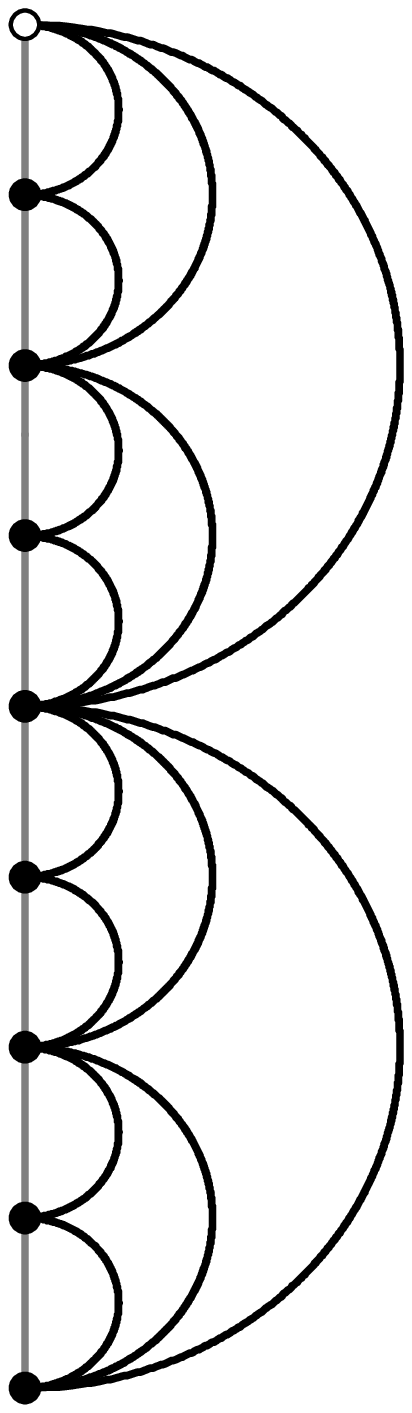}
\end{center}
\vspace{-5mm}
\caption{\label{fig:graph}
Recursive construction of a shortcut network.
The number of generation, $n$, equals 1,2,3 and 4 from left to right. 
The vertical lines and solid arcs indicate backbone edges 
and shortcut edges, respectively.
A periodic boundary condition is imposed in the vertical direction. 
The graph with $n=1$ is a single node with a self-connecting edge. 
}
\end{figure}


Next, we examine the validity of the present scaling ansatz 
by considering a hierarchical small-world network 
that is constructed in a recursive manner  
as shown in Fig.~\ref{fig:graph}. 
The graph with $n$ generations has $N = 2^{n-1}$ nodes 
and $3 \times 2^{n-1}-2$ edges.
The degree distribution function is exponentially decaying 
as $P_k \propto 2^{-k/2}$. 
We note the edges represented by the vertical lines in Fig.~\ref{fig:graph} 
as backbone edges (BBEs) and arcs as shortcut edges (SCEs). 

The energy function of the $q$-state Potts model on the network 
under magnetic field $H$ is 
\begin{eqnarray}
-E/k_\mrm{B} T = &
\sum_\mrm{\langle i,j \rangle \in BBE} \left[  K + D \deli0 \right] \delij
\nonumber \\ 
& + \sum_\mrm{\langle i,j \rangle \in SCE} \Ell \delij
+ \sum_i H \deli0 , 
\end{eqnarray}
where $\delta_{ab}$ is the Kronecker's delta 
and $\sigma_i$ is a spin variable at $i$-th site 
taking one of the values, $0, 1, \cdots, q-1$.
The first summation is over BBEs and the second one is over SCEs.
We consider only the case $D=0$, 
but $D$ becomes finite  in the real space RG performed below.
In the following we consider the case with $q \ge 3$
\footnote{
The present model with $q=2$ 
exhibits a behavior qualitatively different from those with $q\ge 3$,
which is characterized by a pitchfork bifurcation of fixed points 
and will be discussed elsewhere.
The percolation model ($q=1$) on a similar network 
investigated in Ref.~\cite{ Boettcher12} also shows a pitchfork bifurcation.
}
.

\begin{figure}[t]
\begin{center}
\includegraphics[trim=40 60 80 20,scale=0.32,clip]{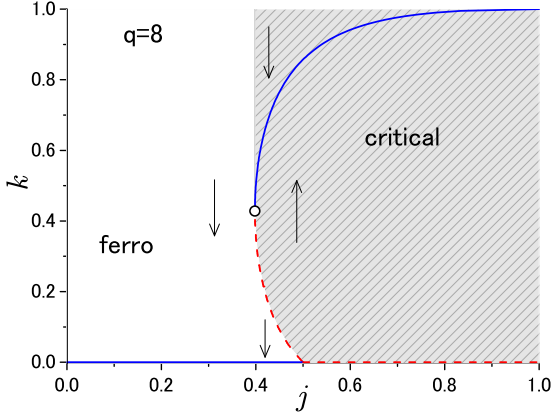}
\end{center}
\vspace{-5mm}
\caption{\label{fig:PD}
(color online) 
Phase diagram for $q=8$.
The shaded region indicates the critical phase. 
The (blue) solid line and the (red) dashed line 
denote the stable and unstable fixed lines, respectively. 
The circle symbol denotes the SNB point, $(\ell\sn, k\sn)$. 
}
\end{figure}


We can calculate the partition function of this system in a stepwise manner; 
decimating the spins in the youngest generation, 
which is the inverse procedure to grow the graph. 
The partial sum of the partition function preserves the function form 
by replacing the parameters of the energy function as  
\begin{eqnarray}
C\n+ e^{
( K\n+ + D\n+ \deli0 ) \delij + (H\n+/2) (\deli0 + \delj0)
}
\nonumber \\
= C_n^2 e^{
(H_n/2) ( \deli0 + \delj0 )
}
\nonumber \\
\times
\sum_{\sigma_k=1}^{q} 
e^{
( K_n + D_n \delta_{\sigma_k 0} + \Ell )
( \delta_{\sigma_k \sigma_i} + \delta_{\sigma_k \sigma_j} ) 
+ H_n \delta_{\sigma_k 0} 
}.
\end{eqnarray}
where $\sigma_i$ and $\sigma_j$ are nearest spins 
that are older than $\sigma_k$ by one generation. 
This gives us the following recursion relations: 
\begin{eqnarray}
g\n+ &=& g_{n} + A^{(4)}_n/2^{n+1} 
\label{eq:eq_of_g}
\\
K\n+ &=& A^{(2)}_n - A^{(4)}_n 
\label{eq:eq_of_K}
\\
H\n+ &=& H_n + 2( A^{(3)}_n - A^{(4)}_n) 
\label{eq:eq_of_H}
\\
D\n+ &=& A^{(1)}_n - A^{(2)}_n - 2 (A^{(3)}_n - A^{(4)}_n)  
\label{eq:eq_of_D}
\end{eqnarray}
with $g_0=0$, $K_0=K$, $H_0=H$, and $D_0=0$. Here
\begin{eqnarray}
e^{A^{(1)}_n} &\equiv& e^{2(K_n+D_n+\Ell)+H_n} + q_1 
\\
e^{A^{(2)}_n} &\equiv& e^{2(K_n+\Ell)} + e^{H_n}+ q_2 
\\
e^{A^{(3)}_n} &\equiv& e^{K_n+D_n+H_n+\Ell} + e^{K_n+\Ell} + q_2 
\\
e^{A^{(4)}_n} &\equiv& 2e^{K_n+\Ell} + e^{H_n} + q_3.  
\label{eq:recursion}
\end{eqnarray}
and $q_m=q-m$.
The quantity $g_n \equiv 2^{-n} \ln C_n$ is regarded as the free energy per spin 
of the system with $n$ generations. 
Note that $\Ell$ does not change in this procedure.

In case of no magnetic field, $H=0$, 
both $H_n$ and $D_n$ remain zero and 
Eq.~(\ref{eq:eq_of_K}) is rewritten as 
\begin{equation}
k\n+ = \frac{ \ell k_n (q_2 \ell k_n + 2 ) }{ q_1 \ell^2 k_n^2 + 1 }, 
\label{eq:recursion_of_k}
\end{equation}
where we put $k_n \equiv e^{-K_n}$ and $\ell \equiv e^{-\Ell}$. 
The fixed point (FP) is obtained from $k_{n+1}=k_n=k^*$ as 
\begin{equation}
k^* \equiv 0, \frac{q_2}{2q_1} 
\left[1 \pm \sqrt{1+ \frac{4q_1}{q_2^2}\frac{2\ell-1}{\ell^2} } \ \right].
\end{equation}
Figure~\ref{fig:PD} shows the RG fixed point 
and the phase boundary in $k$ vs $\ell$ space for $q=8$.
This system exhibits a phase transition 
from the ferromagnetic phase corresponding to the FPs with $K=\infty$ 
to the phase corresponding to the FPs with finite $K$ 
by increasing $\ell$ at fixed $k$. 
We call the latter `critical phase' in the sense that 
the RG flow goes to neither $K=0$ nor $K=\infty$ 
but to nontrivial fixed points. 
The FP exhibits a saddle-node bifurcation (SNB) located at 
\begin{eqnarray}
(\ell\sn, k\sn) \equiv  
\left( 
 \frac{4q_1}{q_2^2} \left[ 
  \sqrt{1+ \frac{q_2^2}{4q_1} } - 1
 \right], 
\frac{q_2}{2q_1}
\right)
\nonumber
\end{eqnarray}
For $k<k\sn$, the phase boundary is given by a line consisting of the unstable FPs, 
each of which leads to a PLS. 
On the other hand, a transition with ES occurs at $\ell=\ell\sn$ for $k \ge k\sn$ 
and its singularity is governed by the SNB point. 
In the following, we consider the phase transition 
for $k=k\sn$ in increasing $\ell$ (decreasing $\Ell$).



First, we perform linear instability analysis around the SNB point. 
Equations~(\ref{eq:eq_of_K})-(\ref{eq:eq_of_D}) are approximated as 
\begin{eqnarray}
\left(
\begin{array}{c}
K_{n+1} \\ D_{n+1} \\ H_{n+1}
\end{array}
\right)
= \mrm{M} 
\left(
\begin{array}{c}
K_{n} \\ D_{n} \\ H_{n}
\end{array}
\right), 
\end{eqnarray}
\begin{eqnarray}
\mrm{M} \equiv
\left(
\begin{array}{ccc}
1 & 0 & y(P_1y-P_2) \\
0 & 2(P_1-P_2) & (1-y)[P_1(1+y)-2P_2] \\
0 & 2P_2 & 2P_2(1-y)+1 
\end{array}
\right) ,
\end{eqnarray}
where $y=\ell\sn k\sn$, $P_1=1/(1+q_1 y^2)$ and $P_2=1/(2+q_2 y)$.
The eigenvalues of the matrix M are 1 and 
\begin{eqnarray}
\lambda_\pm \equiv 1 + P_3 \pm \sqrt{P_3^2 - 2P_2(1-y)[P_1(1-y)-1]}, 
\nonumber
\end{eqnarray}
with $P_3 \equiv P_1-P_2 y-1/2$. 
The largest eigenvalue is $\lambda_+$, and thus 
$H_n$ grows as $e^{n y_h}$ with $y_h = \ln \lambda_+$.

For $H=0$, only $\lambda=1$ is the relevant eigenvalue, 
which means that the instability is marginal as expected at the SNB point.
Equation~(\ref{eq:recursion_of_k}) is rewritten as 
\begin{eqnarray}
s_{n+1} - s_n = A ( s_n^2 + B t)
\label{eq:Kosterlitz_eq}
\end{eqnarray}
in the lowest order of $s_n$ and $t$, where 
$s_n = k\sn - k_n$, $t = \ell\sn - \ell$, 
$A \equiv q_1 \ell\sn^2 k\sn/[q_1 \ell\sn^2 k\sn^2 + 1]$ 
and $B=2(1-\ell\sn)/q_1 \ell\sn^3$.

Equation~(\ref{eq:Kosterlitz_eq}) can be approximated 
to the so-called Kosterlitz equation \cite{Kosterlitz74}, 
$$d s/dn = A [ s^2 + B t ] \nonumber$$ 
which is solved with 
$$
s_n(t) = \sqrt{B t} \tan \left[An \sqrt{B t} \right] 
\nonumber
$$
for the initial condition $s_0=0$.
In this expression [valid until $s_n$ becomes $O(1)$], 
$s_n$ divergently grows as 
\begin{eqnarray}
s_n(t) = 1/A ( n_0(t) - n )
\ \  \mrm{with} \ \ 
n_0(t) \equiv \pi/2A\sqrt{Bt}.
\label{eq:s-n-t}
\end{eqnarray}

Now let us consider the scaling behavior of the present model. 
From the aforementioned analysis, the free energy behaves as 
\begin{eqnarray}
g(s=0, t, H, N^{-1}) = 2^{-n} 
g \left( s_n(t), t, H e^{y_h n}, N^{-1} 2^{n} \right) . 
\nonumber
\end{eqnarray}
When we assume that the second argument $t$ does not yield any singularity, 
we obtain Eq.~(\ref{eq:free_energy2}) by considering that 
\begin{eqnarray}
b=2^n, \ \ 
y_h = \ln \lambda_+, \ \ 
x_t = \frac{1}{2} \ \ 
\mrm{and} \ \  
t_0 = \frac{1}{B} \left( \frac{\pi \ln 2}{2A} \right)^2, 
\nonumber
\end{eqnarray}
where we use Eq.~(\ref{eq:s-n-t}) and $2^{n_0(t)-n}=\xi(t) b^{-1}$.
The value $x_t=1/2$ must have certain universality, 
because it is directly derived from the simplest nonlinear RG equation, 
$d s/dn \propto s^2 + B t$.
This equation is a consequence of the SNB at the edge of the stable fixed line. 
We consider that this structure of the RG flow is the essence of the ES. 
In fact, it has been found in some systems 
\cite{Hinczewski06, Berker09, Hasegawa10, Boettcher11}.


Finally we confirm the scaling ansatz 
by calculating the order parameter, 
$m_n = ( \tilde{m}_n - q^{-1})/(1-q^{-1})$ 
where $\tilde{m}_n = \del[g_{n-1} + {Z_{n-1}}]/\del H$ 
and $Z_n = e^{2(K_n+\Ell+D_n+H_n)} + q_1 [ e^{2(K_n+\Ell)} + 2e^{H_n} + q_2 ]$.
For this aim, we additionally calculate 
the derivatives of $g_n, K_n, D_n$ and $H_n$ with respect to $H$,  
whose recursion equations for these quantities 
are obtained by differentiating Eqs.~(\ref{eq:eq_of_g})-(\ref{eq:eq_of_D}). 
Similarly, we calculate the susceptibility by using the second derivatives.

\begin{figure}[t]
\begin{flushleft}
\hspace{.6cm} {\large \bf (a)}\\
\end{flushleft}
\vspace{-1.2cm}
\begin{center}
\includegraphics[trim=30 20 160 30,scale=0.32,clip]{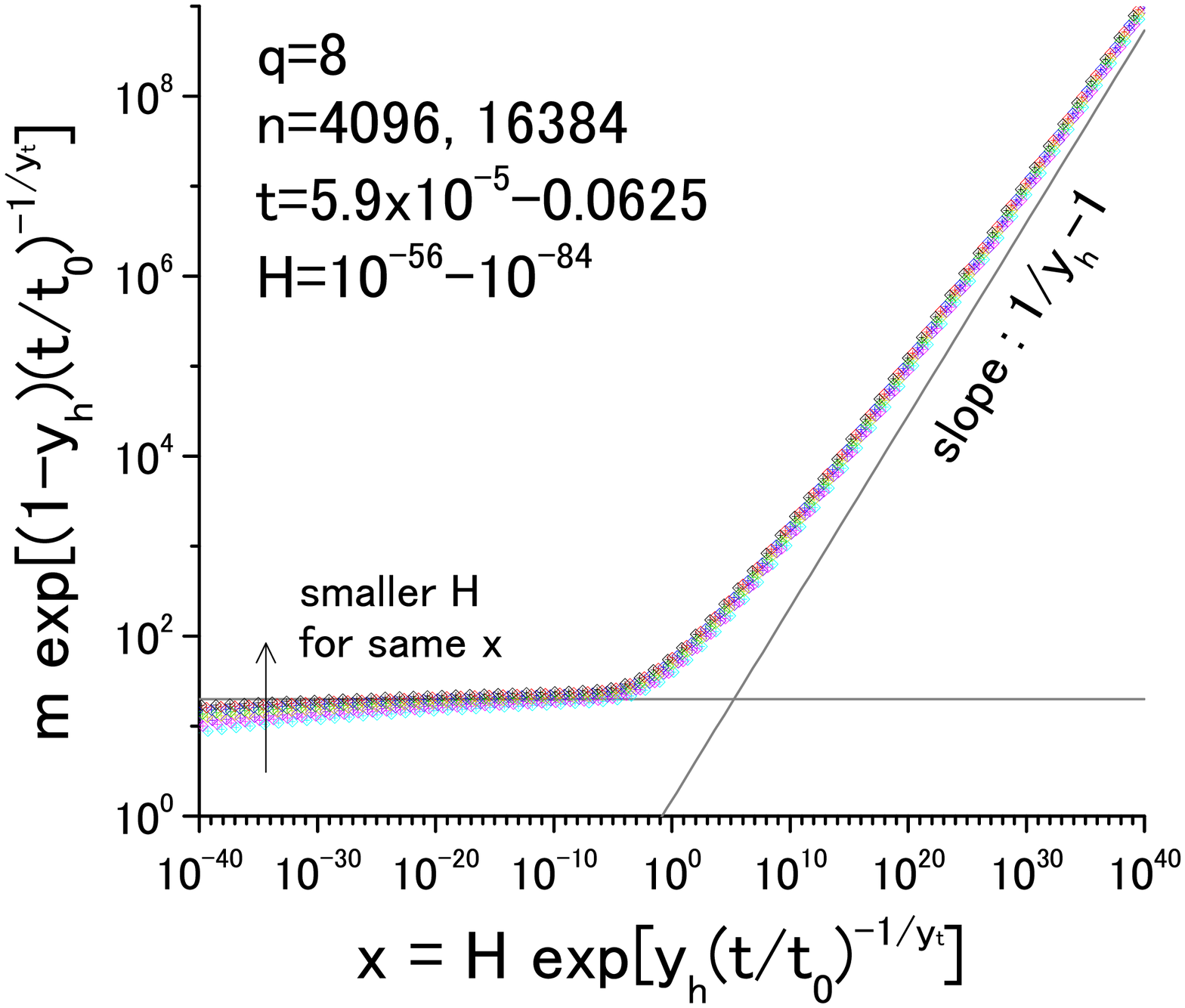}
\\
\end{center}
\vspace{-0.5cm}
\begin{flushleft}
\hspace{.6cm} {\large \bf (b)}\\
\end{flushleft}
\vspace{-1.2cm}
\begin{center}
\includegraphics[trim=30 40 160 20,scale=0.32,clip]{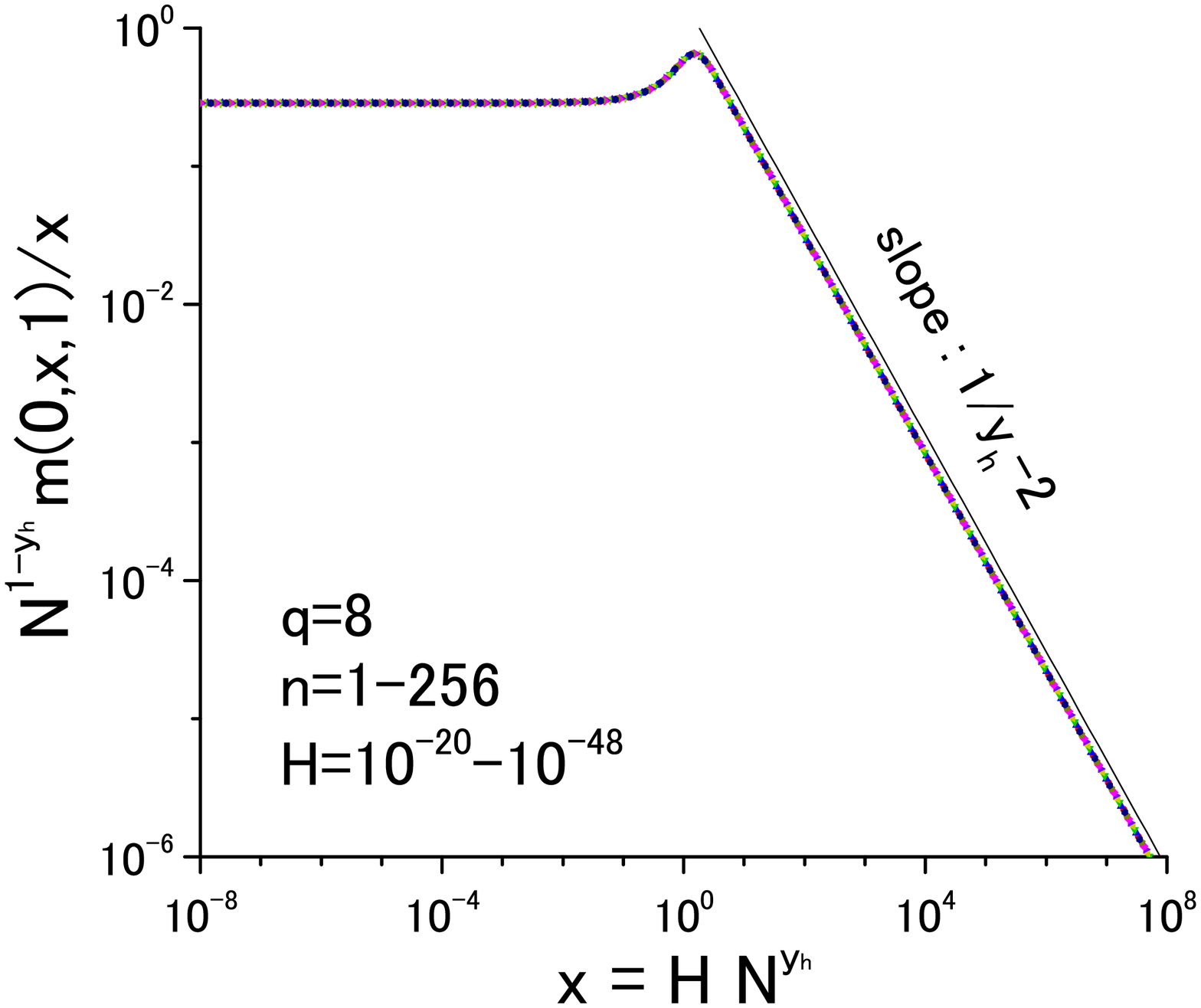}
\end{center}
\vspace{-5mm}
\caption{\label{fig:scaling}
(color online) 
Scaling plot of the order parameter for $1/N \approx 0$ (a) 
and for $t=0$ (b).
In the latter case, we show the data for $n$=4096 and 16384 
to confirm the convergence to the large size limit. 
}
\end{figure}

Figure~\ref{fig:scaling}(a) shows the scaling plot at $N \rightarrow \infty$ 
corresponding to Eqs.~(\ref{eq:scl1a}) and (\ref{eq:scl1b}). 
Although we found a little correction to scaling 
that tends to disappear for $H \rightarrow 0$,  
a good collapse of data is obtained without any fitting parameter.
In Fig.~\ref{fig:scaling}(b), 
we can see excellent scaling behavior for $t=\ell-\ell\sn=0$ 
corresponding to Eqs.~(\ref{eq:scl3a}) and (\ref{eq:scl3b}). 
In both cases, the asymptotic form agrees with the prediction. 




In conclusion, we have proposed a new scaling theory for the ES 
in infinite dimensional systems, 
and clarified its validity by analyzing a simple model. 
We believe that the present scaling formulae can be applied to various other models, and will clarify the existence of the universal mechanism for them. 
We have already confirmed that the scaling law holds in the bond percolation model 
on the decorated (2,2)-flower \cite{Berker09, Hasegawa10}  (not shown here)
and random growing network \cite{Hasegawa10b}.
The finite size scaling formula included in the present theory 
will be useful in analyzing real-world data or numerical simulations, 
in which we can treat only small generations, $n \sim \log N$.



\end{document}